\documentclass[letterpaper, 10 pt, conference]{ieeeconf}  

\pdfoutput=1  

\IEEEoverridecommandlockouts                              

\usepackage{graphics} 
\usepackage{amsmath} 
\usepackage{amssymb} 
\usepackage[compress]{cite}
\usepackage[usenames,dvipsnames]{xcolor}
\usepackage{graphicx}
\usepackage{amsfonts}
\usepackage{mathtools}
\usepackage[mathcal]{euscript}
\usepackage{hyperref}
\usepackage{dblfloatfix}
\usepackage{bm}

\hypersetup{
	colorlinks   = true, 
	urlcolor     = blue, 
	linkcolor    = black, 
	citecolor   = black 
}

\usepackage[T1]{fontenc} 
\usepackage[hyphenbreaks]{breakurl} 
\usepackage{mathtools, cuted} 
\usepackage[ruled,vlined]{algorithm2e} 
\usepackage{enumerate}  
\usepackage{balance} 
\usepackage{tikz}

\usepackage{subcaption}
\DeclareCaptionLabelSeparator{periodspace}{.\quad}
\usepackage{amsthm}
\usepackage{url}

\usepackage[colorinlistoftodos,prependcaption,textsize=tiny]{todonotes}
\usepackage[authormarkuptext=name,addedmarkup=bf,authormarkupposition=left]{changes}

\usepackage[english]{babel}
\usepackage{blindtext}

\theoremstyle{definition}

\newcommand{\norm}[1]{\left\lVert#1\right\rVert}

\graphicspath{{figs/}}

\title{\LARGE \bf
Dynamic Landing of an Autonomous Quadrotor\\%
on a Moving Platform in Turbulent Wind Conditions
}

\author{Aleix Paris, Brett T. Lopez, and Jonathan P. How
\thanks{A.~Paris, B.~T.~Lopez, J.~P.~How are with the Aerospace Controls Laboratory,
        MIT, 77 Massachusetts Ave., Cambridge, MA, USA
        {\tt\small \{aleix, btlopez, jhow\}@mit.edu}}%
}

\begin{document}

\maketitle
\thispagestyle{plain}
\pagestyle{plain}



\begin{abstract}
Autonomous landing on a moving platform presents unique challenges for multirotor vehicles, including the need to accurately localize the platform, fast trajectory planning, and precise/robust control. Previous works studied this problem but most lack explicit consideration of the wind disturbance, which typically leads to slow descents onto the platform.
This work presents a fully autonomous vision-based system that addresses these limitations by tightly coupling the localization, planning, and control, thereby enabling fast and accurate landing on a moving platform.
The platform's position, orientation, and velocity are estimated by an extended Kalman filter using simulated GPS measurements when the quadrotor-platform distance is large, and by a visual fiducial system when the platform is nearby.
The landing trajectory is computed online using receding horizon control and is followed by a boundary layer sliding controller that provides tracking performance guarantees in the presence of unknown, but bounded, disturbances.
To improve the performance, the characteristics of the turbulent conditions are accounted for in the controller. The landing trajectory is fast, direct, and does not require hovering over the platform, as is typical of most state-of-the-art approaches. Simulations and hardware experiments are presented to validate the robustness of the approach.
\end{abstract}


\section*{Supplementary Material}

Video of the paper summary and experiments is available at \url{https://youtu.be/xKo1rY4riJQ}.


\section{INTRODUCTION}\label{s:intro}
Autonomous unmanned aerial vehicles (UAVs) are becoming more and more popular in industry for their flexibility and fast deployment, and have demonstrated their usefulness in applications such as aerial photography for topology and agriculture \cite{mancini2013using, anderson2013lightweight, honkavaara2013processing, tokekar2016sensor}, search and rescue operations \cite{tomic2012toward, scherer2015autonomous}, and mapping \cite{caballero2009vision, schmuck2017multi, vidal2018ultimate}. 
The large recent growth  in online shopping has also attracted interest in reducing package shipment time and costs, and UAVs provide an efficient alternative to delivery trucks\cite{scott2017drone, shavarani2018application}.
However, their payload and flight time is limited,
so several researchers have investigated using a truck-drone delivery system \cite{ham2018integrated, ferrandez2016optimization}.
The UAVs in current truck-drone delivery methods can only take off and land when the truck is stopped visiting a customer node, which has substantial synchronization costs. Thus, autonomous landing on a moving truck is a promising research direction~\cite{otto2018optimization}.
This work presents a system capable of landing a quadrotor on a moving platform, even in the presence of turbulent wind.
We demonstrate a boundary layer sliding controller (BLSC) which takes into account these conditions, a planner with changing objectives that allows a fast maneuver, and a vision-based extended Kalman filter (EKF) to estimate the moving platform's state.
The UAV's state estimator currently uses a motion capture system, but the important information in the maneuver we demonstrate (the relative position of the UAV and the moving platform) is estimated onboard by the quadrotor using a vision system.

\begin{figure} [t!]
	\begin{center}
		\includegraphics[trim =25mm 70mm 60mm 55mm, clip, width=0.4\textwidth]{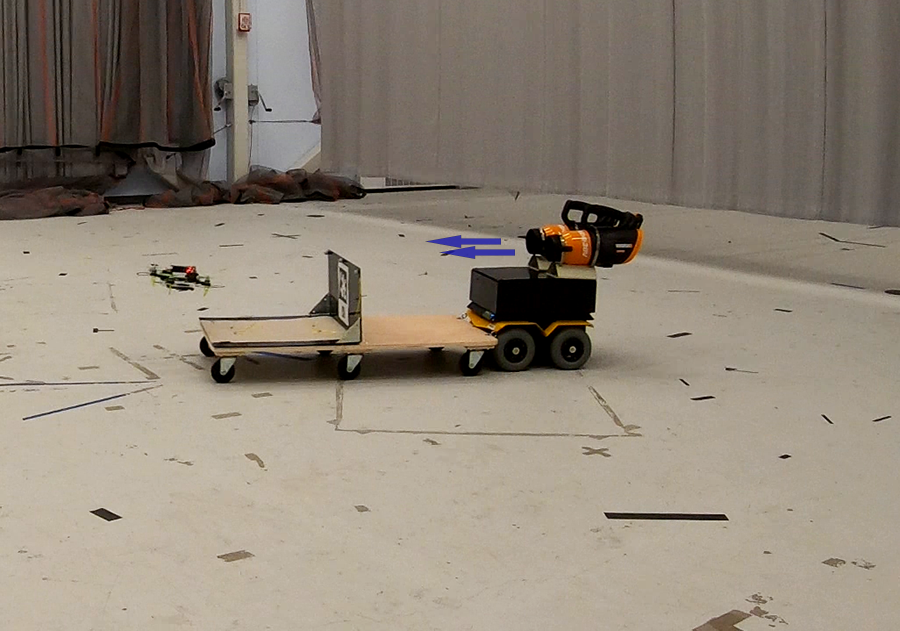}
		\caption{Dynamic landing maneuver on a moving platform in a wind environment generated by two leaf blowers.}
		\label{fig:video-screenshot}
	\end{center}
\end{figure}

\subsection{Related Work}

Autonomous UAV landing has been investigated by several researchers.
Ref.~\cite{venugopalan2012autonomous} landed a drone on a static kayak in a reservoir with mild wind and water ripple conditions, but the image processing was done off-board and the landing time was close to 1~min.
Ref.~\cite{araar2017vision} developed a system capable of landing a UAV on a moving platform with PID control, an EKF estimator, and an AprilTag visual fiducial bundle.
However, the computation was also done off-board, and there was no relative wind present: the platform's speed was just 0.18m/s and the tests were done indoors.
Ref.~\cite{falanga2017vision} demonstrated a quadrotor landing on a moving platform using only onboard sensing and computation, but the environment was not turbulent due to the tests being indoors and the platform was moving relatively slowly at 1.2m/s.
Furthermore, besides two cameras, a distance sensor was required to estimate the UAV-platform relative pose.
Ref.~\cite{borowczyk2017autonomous} demonstrated an autonomous landing of a quadrotor on a car moving at 14m/s by detecting an AprilTag on its roof.
A proportional navigation-based guidance law was used for the approach phase, and a PID controller for the landing phase, with no disturbance rejection considerations. The landing maneuver consisted of acquiring the tag while hovering, and the descent was initiated once the quadrotor stabilized over it. Additionally, the quadrotor used was large and had a broad sensor suite, including a downward-facing camera, a three-axis gimballed camera to track the target, and an inertial navigation system. Furthermore, besides the ground vehicle's GPS coordinates, the quadrotor also used the ground vehicle's IMU data to improve its estimation of the landing platform's state.
Ref.~\cite{xing2019autonomous} also demonstrated an outdoors landing, but the landing platform's speed was only 0.5m/s and the landing maneuver took 12--20~s to complete. Moreover, the ground vehicle's wheel encoders data was used to estimate its state.
Ref.~\cite{persson2019model} used model predictive control (MPC) to land a quadrotor on a moving ship. Both the UAV and the ship collaborated to reach their goal.
The waves were modeled as sinusoidal, but the wind disturbances considered were not turbulent -- this approach only compensated the effects caused by a steady-state wind. Furthermore, the vehicle needed to hover above the platform for at least 5~s, and the ship was simulated.
A simulated boat landing was also carried out in \cite{wynn2019visual}, where the platform's state was estimated fusing GPS and visual measurements.
This work incorporated a velocity feed-forward term to the controller to catch the platform, but its only consideration of external wind was an offset of the hovering position to ensure the target was inside the field of view of the downwards-facing camera, and the landing maneuver lasted 24~s.
Ref.~\cite{bhargavapuri2019vision} developed an adaptive controller to track a ground vehicle with only relative position data from ArUco tags, and tested it in outdoor experiments at 5.6m/s. But this approach only considered disturbances due to ground effect, and the landing maneuver needed 20~s for tracking and 10~s for descent.

In summary, most current approaches for quadrotor landing on moving platforms involve hovering above the platform for a period of time to visually acquire it, which is then followed by a relatively slow descent.
In contrast, our approach investigates a direct trajectory to the landing platform, which has the possible advantages of providing faster landings and enabling several UAVs to approach the platform at the same time from different directions (improving the utilization of this limited resource).
Additionally, most of current approaches do not make special considerations to reject the turbulent wind present near the target vehicle, and thus safety in challenging conditions cannot be guaranteed.

\subsection{Contributions}

 This paper demonstrates a vision-based system capable of dynamic landing (i.e., the multirotor does not need to hover above the vehicle before descending) which also accounts for the turbulent conditions that would be present near a rapidly-moving ground vehicle. The resulting framework allows thus a maneuver that will be crucial for efficient truck-drone delivery systems. The contributions of this paper are as follows:
\begin{itemize}
    \item Demonstration of a boundary layer sliding controller to incorporate and compensate for turbulence based on a model of the conditions near the landing platform.
    \item An algorithm for computing fast, vision-based dynamic landing maneuvers, is demonstrated both in simulation and hardware experiments that include challenging steady/turbulent wind conditions.
\end{itemize}

\section{SYSTEM OVERVIEW}\label{s:system}

\begin{figure} [t]
	\begin{center}
		\includegraphics[trim =5mm 70mm 5mm 35mm, clip, width=0.47\textwidth]{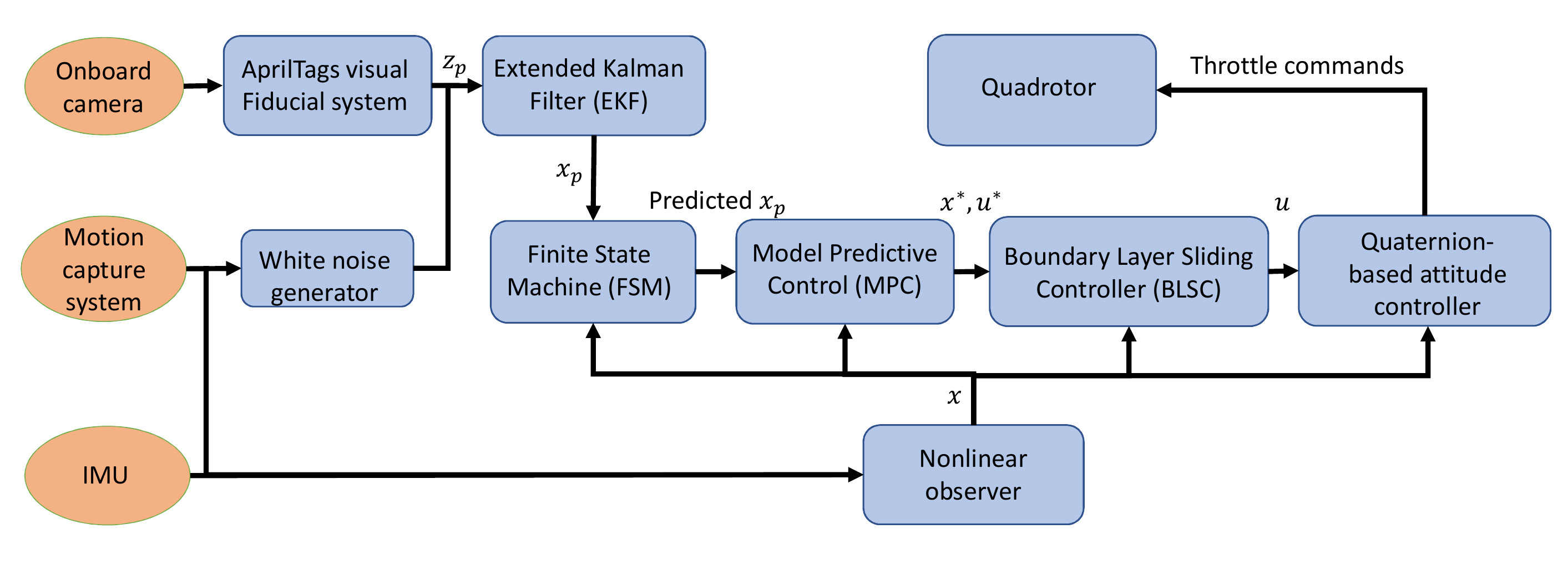}
		\caption{Diagram of the system's architecture. Orange circles represent the inputs to the system, and blue squares represent each component. The arrow's label indicates the information that is sent.}
		\label{fig:system-diagram}
	\end{center}
\end{figure}

This work addresses the current limitations of landing on a moving platform by: 1) using optimization-based trajectory generation to enable dynamic landing; and 2) using robust control to explicitly compensate for turbulent wind conditions. The system that achieves this is comprised of several components, described in this section and shown in Fig.~\ref{fig:system-diagram}.

\subsection{Finite State Machine}\label{subs:fsm}
The quadrotor's behavior is determined by a finite state machine (FSM) comprised of four states:
\begin{enumerate}
    \item \textit{Stand By}: This is the initial state of the quadrotor, which consists of taking off and hovering at a predefined altitude above the starting point. The FSM then transitions to \textit{Search} mode.
    \item \textit{Search}: The quadrotor uses simulated GPS coordinates of the unmanned ground vehicle (UGV) ---as explained in Section~\ref{subs:estimation}--- to predict a rendezvous location and flies there. When the front-facing camera detects the landing platform as described in Section~\ref{subs:detection}, the state automatically switches to \textit{Landing}.
    \item \textit{Landing}: In this mode, the quadrotor approaches the target following a direct trajectory towards it. When the distance and relative velocity UAV-UGV are below threshold values, the quadrotor switches to the \textit{End} mode. If the last detection happened more than 0.8s ago, the mode returns to \textit{Search}.
    \item \textit{End}: motors are stopped, maneuver is finished.
\end{enumerate}
\subsection{Trajectory Planning: Model Predictive Control}\label{subs:planning}

The planner solves a convex optimization problem with changing objectives depending on the state of the FSM.
In \textit{Search} mode, the UAV-UGV rendezvous point is predicted by assuming a constant linear velocity and yaw rate for the UGV, and a constant velocity for the UAV.
This position is then offset by a small distance backwards in the direction of the UGV to ensure target detection by the front-facing camera, and is used by the planner as the final position of the trajectory, while the final velocity is the UGV's.
The time taken to reach the UGV is minimized, to reduce delivery turnarounds.
In \textit{Landing} mode, the planner initially minimizes the jerk to obtain a trajectory which ensures adequate tag acquisition.
As the UAV approaches the target, the effect of disturbances increases so the planner minimizes the time spent in the turbulent area. The final position of the planned trajectory is the vertical tag's position (offset a few cm backwards) and predicted ahead (by an amount that depends on the computation time) assuming constant linear/angular velocities.
The final velocity of the trajectory is set to match the UGV's.

The minimum jerk approach produces smooth trajectories and has a long heritage for planning quadrotor paths \cite{yu2014minimum, phang2015systems, falanga2017vision}.
The trajectory is re-planned using an MPC approach every time a new estimate of the platform's position is obtained. We used CVXGEN \cite{MattingleySB12} to generate the code to solve the following convex optimization problem
\begin{equation}
\begin{aligned}\label{eq:opt_problem}
    \min_{x,u}  J &= \sum_{t=0}^{N} \ell \left( \bar{u}, d \right)[t] \\ 
    \text { subject to } x[t+1] &= Ax[t] + B\bar{u}[t] \\
    |a[t] |_{\infty} & \leq a_{\max} , \text{ and } |\bar{u}[t]|_{\infty} \leq j_{\max}   \\
    x[0] &=x_{0}, \, x[N]=x_f  \\
    \text {for } t &= 0\ldots N  
\end{aligned}
\end{equation}
where $N$ is the timestep when the quadrotor has to reach the final state $x_f$, $\ell$ is a quadratic cost function, $\bar{u}$ is the open-loop control input (the jerk of the trajectory), $d$ is the UAV-tag distance, $x$ is the position, velocity, and acceleration of the UAV, $A$ and $B$ are, respectively, the state and input matrices for a triple integrator, and $a$ is the subvector of $x$ representing the acceleration.
Note that we can plan using this linear model for a nonlinear system because the nonlinear dynamics of the quadrotor are canceled by the ancillary controller, as explained next.

\subsection{Ancillary Controller: Boundary Layer Sliding Controller}\label{subs:controller}

While MPC has been used extensively in industry \cite{qin2003survey}, the control of systems with nonlinear dynamics requires expensive optimization. Sliding control \cite{slotine1991applied} has proven to be effective in quadrotors \cite{runcharoon2013sliding, yang2016attitude}. This control strategy guarantees bounds on the tracking error and has been combined with MPC \cite{rubagotti2010robust, lopez2018robust}. In our approach, we derive a nonlinear ancillary controller using sliding control that models the disturbances found near the landing platform. 

In quadrotors, the attitude dynamics are much faster than the position dynamics, and thus control of both can be decoupled \cite{how2015linear}: the output of a controller is the setpoint for the other. The position and velocity controller is derived in this section to account for the turbulent wind present near the landing platform, and the attitude control is performed by a quaternion-based controller \cite{lopez2016low}.

The following derives the BLSC. Define the manifold $S(t)$ by
$    s=\dot{\tilde{x}}+\lambda \tilde{x} = 0,$
where $\tilde{x} = x - x_d$ and $\lambda > 0$. The objective of sliding control is to maintain $s=0$ at all times.
If the control action's frequency is high enough, zero tracking error is guaranteed \cite{slotine1991applied}.
This high control action is impractical in many applications because of actuator limits and the excitation of unmodelled dynamics.
An approach taken in \cite{runcharoon2013sliding, lopez2018robust} is BLSC, where the control discontinuity is smoothed in a thin boundary layer of thickness $\Phi$:
\begin{equation}
    \mathcal{B} := \{s :|s| \leq \Phi\}
\end{equation}

Consider a system whose dynamics can be expressed as
\begin{equation}\label{eq:dynamics}
    \ddot{x} =  f\left(x\right) + b \left( x \right) u + d
\end{equation}
where $d$ is the disturbance. Then, the BLSC strategy is
\begin{equation}\label{eq:u}
    u = \hat{b}^{-1}\left[\ddot{x}_{d}-\lambda \dot{\tilde{x}}-\hat{f}(\boldsymbol{x})-K \operatorname{sat}\left( \frac{s}{\Phi} \right)\right]
\end{equation}
where $\operatorname{sat}(\cdot)$ is the saturation function, $\hat{f}$ is the estimated acceleration caused by drag, and $K$ is determined by the uncertainty in the dynamics and the disturbance of the system. As noted before, we can plan in~\eqref{eq:opt_problem} using linear MPC because of the cancellation of $f$ in~\eqref{eq:dynamics} and~\eqref{eq:u}.

We generate turbulence using leaf blowers as shown in Figs.~\ref{fig:video-screenshot} and~\ref{fig:lbarray}.
The turbulent wind parameters are the mean $\bm{v_w}$ and standard deviation $\sigma$ of the speed. Define
$    \bm{V} = \bm{v} + \bm{v_w}$
where $\bm{v}$ is the quadrotor's speed. Then, $\bm{V}$ is the total wind speed relative to the UAV and $\hat{f}$ is
$    \hat{f} = \hat{c} \norm{\bm{V}} V,$
where $\hat{c}$ is the estimated drag coefficient of the quadrotor.

The quadrotor plans a trajectory to approach the UGV in the direction it is facing to match its speed, and thus is never outside the wind field generated by the leaf blowers during the landing maneuver. Therefore, it is reasonable to assume that this wind field is constant in the directions perpendicular to where the leaf blowers point to. The true acceleration caused by drag is 
\begin{equation}
    f = c \norm{\bm{V} \pm 2\sigma \bm{u_w}} \left( V \pm 2\sigma u_w \right)
\end{equation}
where $\bm{u_w}$ is a unit vector in the direction of the wind.

The variation of $b$ is very small for a UAV with constant weight. Therefore, $\beta = \left(b_{max}/b_{min}\right)$ is approximately 1, where $b_{max}$ and $b_{min}$ are the maximum and minimum control gains respectively (or throttle gains in the context of quadrotors). Thus, $K$ is simplified as \cite{slotine1991applied}
$    K = \bar{F} + \eta,$
where $\eta > 0$ is a constant in the sliding condition
\begin{equation}
    \frac{1}{2} \frac{d}{d t} s^{2} \leq-\eta|s|.
\end{equation}
The larger the $\eta$, the faster the system will reach the sliding surface. Nevertheless, $K$ should only be as large as the disturbance magnitude requires to avoid a high-frequency control signal. $\bar{F}$ is
\begin{equation}
\begin{split}
    F =& | f - \hat{f} | \leq \bar{F}  \\
    \bar{F} =& | \left( \hat{c} + \tilde{c} \right) \norm{\bm{V} \pm 2\sigma \bm{u_w} } (V \pm 2\sigma u_w) - \hat{c} \norm{\bm{V}} V |
\end{split}
\end{equation}
where $\tilde{c} > 0$ is a bound on the absolute value of the drag coefficient error $|c - \hat{c}|$. By taking the sign that makes this coefficient larger, we have defined $K$. In our application, the quadrotor moves towards the generated wind and therefore this occurs when the $2\sigma$ is \textbf{increasing} the magnitude of $\bm{v_w}$.

\subsection{Landing Platform Estimation: Extended Kalman Filter}\label{subs:estimation}

To estimate the state of the moving platform, an extended Kalman filter (EKF) is used. This filtering algorithm minimizes the mean of the squared error and has demonstrated its efficacy in robot localization \cite{kiriy2002three, georgiev2004localization, teslic2011ekf}. The state vector of the platform is
$    \boldsymbol{x_p} = [p_x, p_y, v_p, \theta, \dot{\theta}]^\top,$
where $p_x,p_y$ are the 2D coordinates, $v_p$ is the magnitude of the velocity, $\theta$ is the orientation angle with respect to the $x$-axis, and $\dot{\theta}$ is the angular velocity.
Since real-world roads are mostly horizontal and in particular our experiments were carried on completely flat surfaces, the velocity in the $z$-direction is not estimated.
The moving platform is modeled as a unicycle with dynamics
\begin{equation}
    \boldsymbol{\dot{x}_p}(t)=f_p(\boldsymbol{x_p}(t))+\boldsymbol{w}(t),
\end{equation}
where $\boldsymbol{w}(t)$ is the process noise, assumed to be a white Gaussian noise. We consider a constant linear and angular velocity, and the UGV dynamics are
\begin{equation}
\begin{aligned} 
\dot{p}_{x} &=v_{p} \cos (\theta), && \dot{v}_{p} = 0, \\
\dot{p_{y}} &=v_{p} \sin (\theta), &&
\ddot{\theta} = 0.  
\end{aligned}
\end{equation}

The measurement vector is 
    $\boldsymbol{z_p} = [p_x, p_y, \theta]^\top$, and 
when a measurement is received, the EKF approach is used to perform an update of the estimated state.
The measurements are obtained in two ways.
First, when the quadrotor is far from the platform (that is, in the \textit{Search} state defined in Section~\ref{subs:fsm}), these measurements are obtained by adding a white Gaussian noise to the ground truth pose of the vehicle (obtained from the motion capture data) to simulate inaccurate GPS measurements that a ground vehicle could provide to the UAV for the rendezvous.
Note that receiving $\theta$ is not necessary to estimate the orientation of a moving platform because its motion could be used to infer that quantity.
Nevertheless, $\theta$ measurements are used in this work, which enables testing for static platform experiments. The update frequency is 2Hz, which is realistic for UAV applications \cite{salih2013suitability}. These first set of measurements are simply used to help the UAV locate the ground vehicle, but that information could be estimated without requiring a link between the two vehicles. 

Second, and most importantly, when the quadrotor is near the ground vehicle and detects the onboard tag, we fuse both the simulated GPS and the visual detection measurements to estimate $\boldsymbol{x_p}$. The vision-based detection is explained in the next subsection, and it provides a far more accurate estimate of the position and orientation of the tag/platform. The 2D positions $p_x$ and $p_y$ and the heading angle $\theta$ are then used to update the platform's state, and the estimated velocity $v_p$ is incorporated into the vector $x_f$ in~\eqref{eq:opt_problem} to ensure the quadrotor matches the moving platform's speed at the landing point.

\subsection{Visual detection: AprilTag visual fiducial system}\label{subs:detection}
When the UAV is relatively close to the platform, visual estimation provides more accurate UAV-UGV poses than GPS. We used the AprilTag visual fiducial system \cite{olson2011tags} to obtain them, and a ROS wrapper \cite{malyuta2018guidance} based on AprilTag 2 \cite{wang2016apriltag} to interface with the core detection algorithm.
A tag bundle is a set of several coplanar tags used simultaneously by the visual fiducial system: the algorithm extracts the information of all of them to estimate a single ``bundle pose''. Thus, they are useful when accurate detection is required, which is the case in this paper.
Additionally, by using tags of different sizes, detection at various distances is ensured. We used a tag bundle comprised of a 14$\times$14cm tag on top of a 5$\times$5cm tag.
Despite the relatively small bundle size, it can be detected at a maximum distance of approximately 3.5m, and a minimum distance of about 5cm. The bundle can be seen in Fig.~\ref{fig:lbarray}.

\section{EXPERIMENTAL RESULTS}\label{s:exp}
\subsection{Simulations}
Due to the difficulty of accurately modeling the complex turbulent wind effects on a quadrotor, our simulations consider steady wind but serve to test our finite state machine, planner, controller, and estimator. For space constraints, simulation experiments are not analyzed here, but the video accompanying the paper (linked in the Supplementary Material Section) shows a landing simulation.

\begin{figure} [t]
	\begin{center}
	\vspace*{.1in}
		\includegraphics[trim =0mm 0mm 0mm 0mm, clip, width=0.35\textwidth]{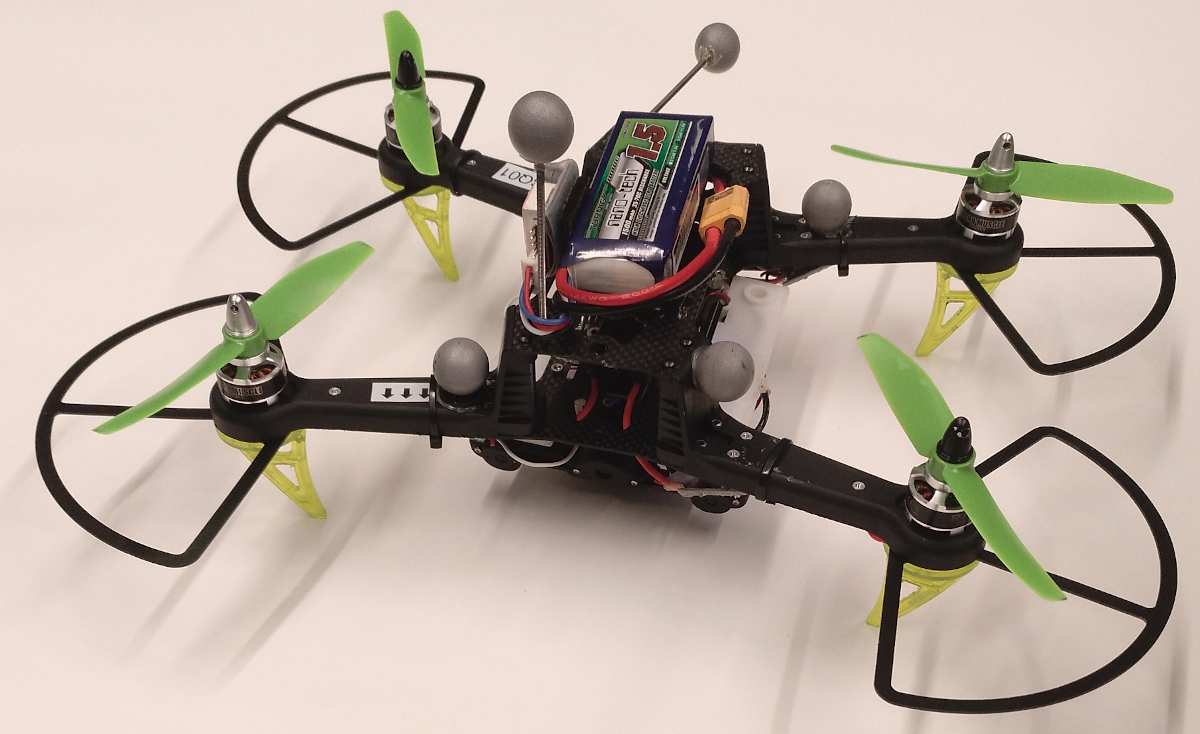}
\vspace*{-.05in}
\caption{Quadrotor vehicle used for the experiments.}
		\label{fig:sq01}
	\end{center}
%
\begin{center}
\includegraphics[trim =10mm 150mm 0mm 0mm, clip, width=0.35\textwidth]{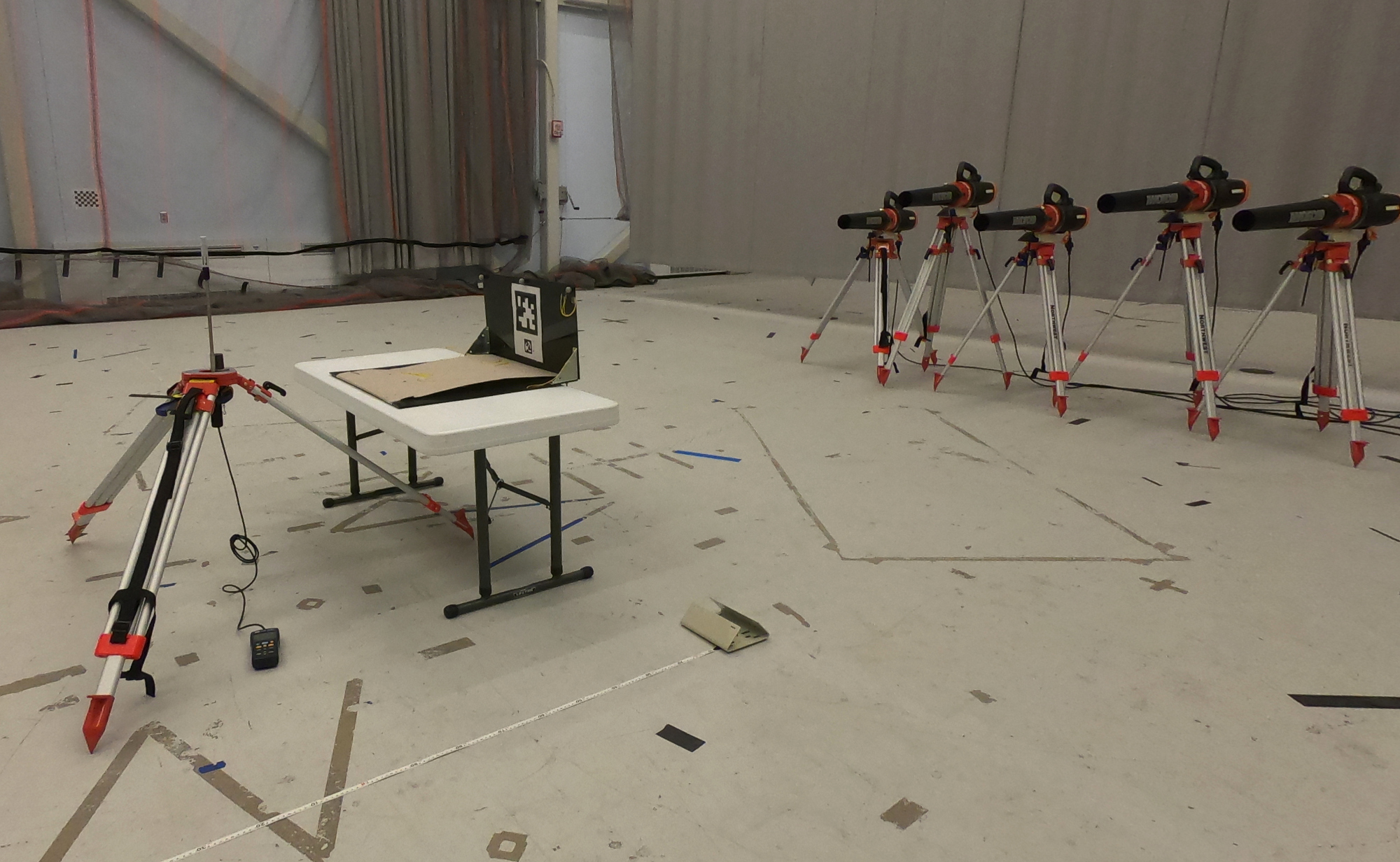}
\vspace*{-.05in}
\caption{Leaf blower array used in front of the static platform showing the wind measurement procedure.}
\label{fig:lbarray}
\end{center}
%
\begin{center}
\includegraphics[trim =0mm 0mm 0mm 0mm, clip, width=0.35\textwidth]{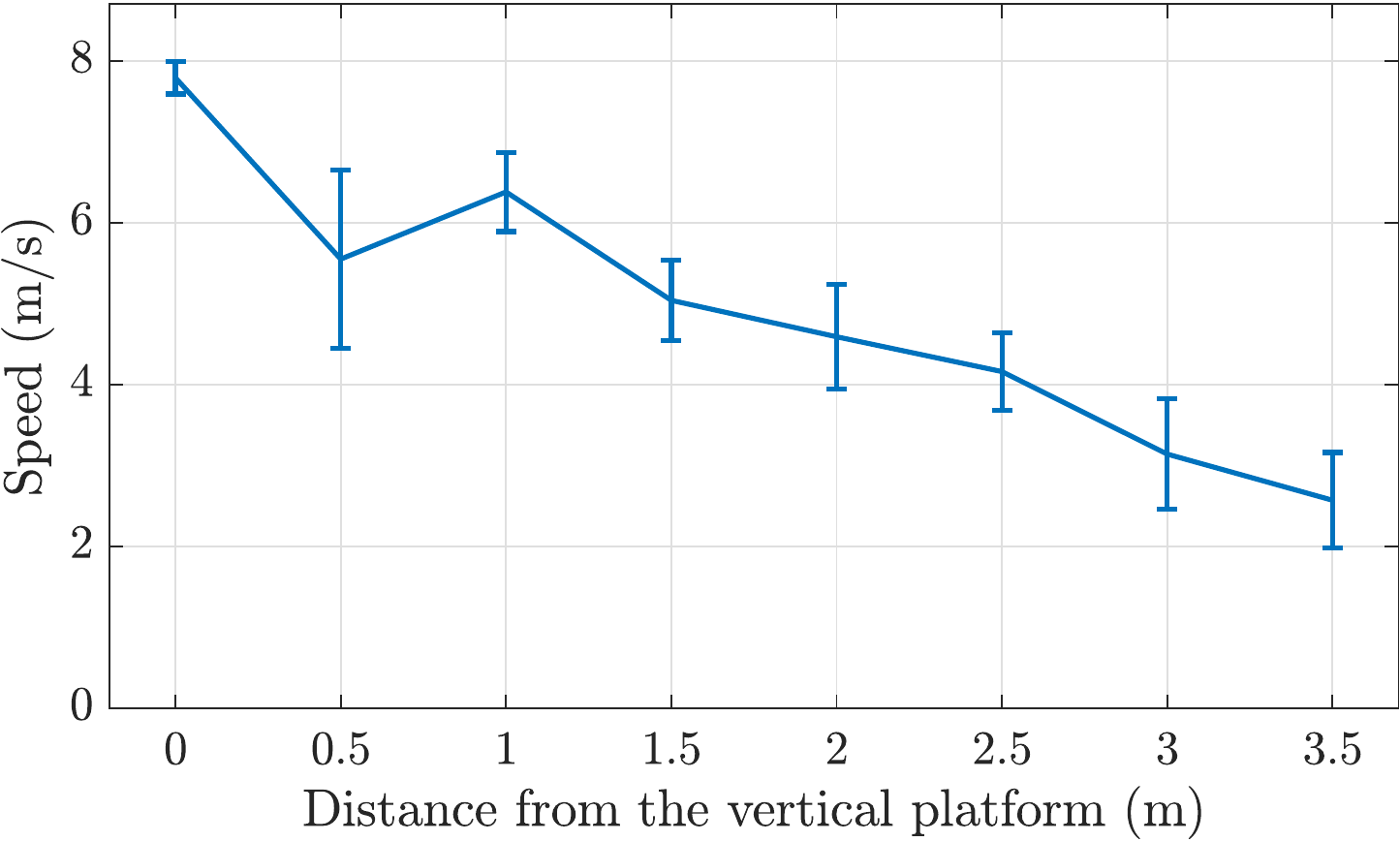}
\vspace*{-.05in}
\caption{Mean wind speed (in m/s) from the leaf blowers vs.~distance to the vertical platform (with $1-\sigma$ error bars).}
\label{fig:windsp}
\end{center}
\end{figure}

\begin{figure*}[ht]
\begin{center}
\vspace*{.05in}
\includegraphics[trim =40mm 5mm 40mm 0mm, clip, width=1\textwidth]{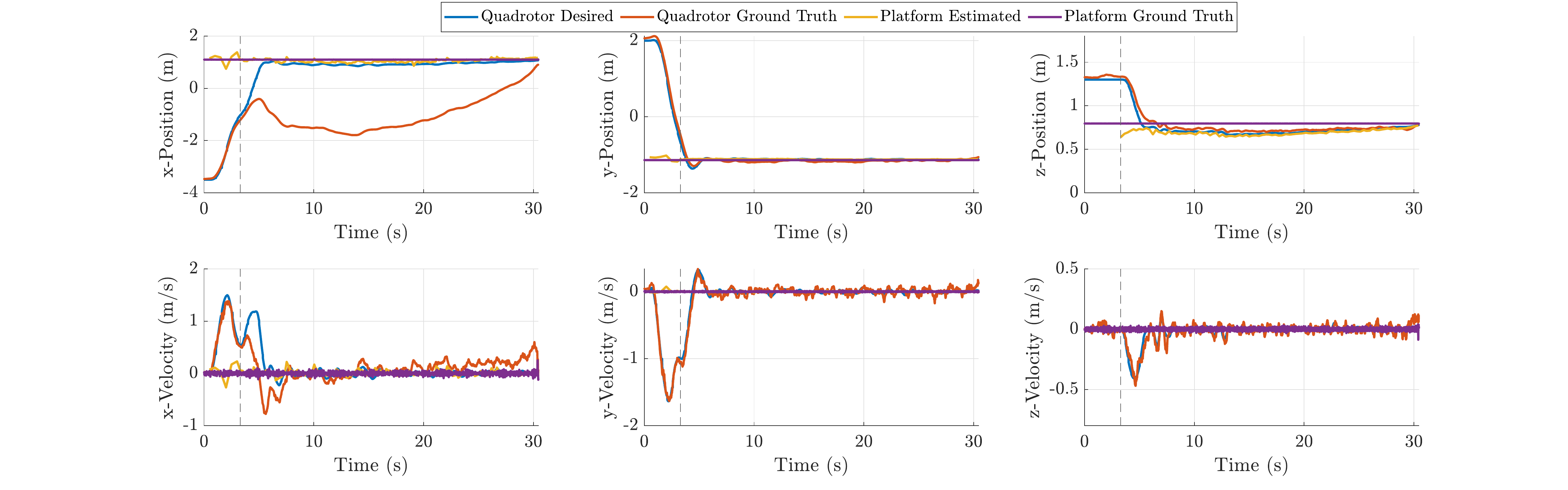}
\vspace*{-.175in}
\caption{Position and velocity tracking and estimation performance of a static platform experiment using a BLSC that does not model the turbulence. Wind blowing towards $-x$ resulting in poor tracking performance in the $x$-direction. Vertical dashed lines indicate time of the first tag detection, $t_d$.}
\label{fig:static_no}
\end{center}
\end{figure*}

\subsection{Hardware}
Static and moving platform landing tests were done (see Figs.~\ref{fig:video-screenshot} and \ref{fig:lbarray}
), both of which had turbulent wind at the landing site from the leaf blowers. The quadrotor used for the hardware experiments (Fig.~\ref{fig:sq01}) weighs 0.564kg including the 1500mAh 3S battery. It is $36\times29$cm (approximately half the platform size) and its thrust-to-weight ratio is 1.75. The onboard computer is a Qualcomm Snapdragon Flight APQ8074, whose front-facing camera provides $640\times480$ black-and-white images at a rate of 30fps to the AprilTag detection module.
Hover tests were carried out to determine $\hat{b}$ in \eqref{eq:u}.
To measure the drag coefficient $\hat{c}$ and the bound on its error $\tilde{c}$, we performed tests with the quadrotor flying in front of a strong wind and the accurate pitch angle was obtained by the motion capture system.
By balancing forces, the drag could be determined, yielding $\hat{c}$.
Additionally, we used the {\tt IMU utils} package \cite{imucal} to compute the Snapdragon's IMU accelerometer and gyroscope noise density and bias random walk.
The {\tt Kalibr} visual-inertial calibration toolbox then used this IMU intrinsic information to find the camera-IMU transform \cite{furgale2015kalibr}.

\subsection{Static Platform Experiments}
\subsubsection{Experiment Setting}
The first set of hardware experiments presents a static platform in front of an array of 5 leaf blowers, as shown in Fig.~\ref{fig:lbarray}. The leaf blowers are set at two different heights and point to the negative $x$-direction.
The platform is composed of a 60$\times$60cm horizontal plate and a 60$\times$30cm vertical plate, which has attached the tag bundle described in \ref{subs:detection}.

The parameters $v_w$ and $\sigma$ of the turbulent wind were measured at distances $l$ from the vertical platform every $0.5$m until $l = 3.5$m, which is approximately the tag detection range. For every $l$, a measurement of the speed was taken every second for $60$s using a high-precision hot-wire anemometer. 
Fig.~\ref{fig:windsp} shows the data obtained. Interestingly, at $l = 0.5$m, the mean speed decreases while the standard deviation increases to its maximum value. We believe this is due to the vortices caused when the flow traverses the vertical platform.

\begin{figure*}[ht]
\begin{center}
\includegraphics[trim =40mm 5mm 40mm 0mm, clip, width=1\textwidth]{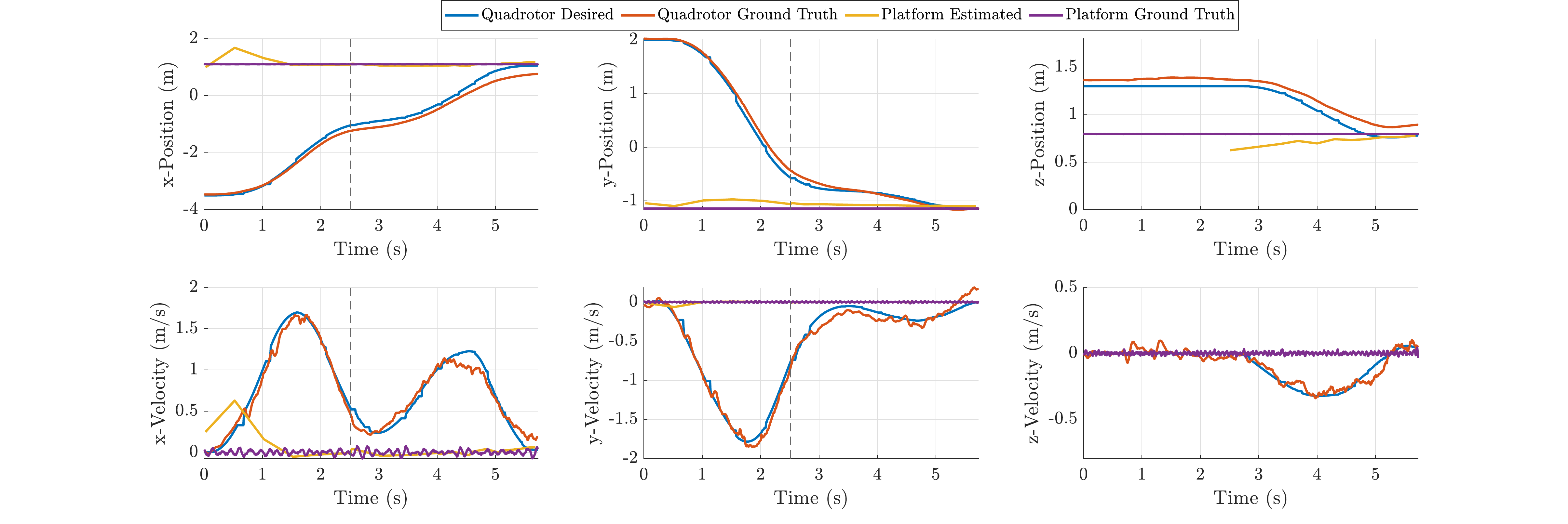}
\vspace*{-.175in}
\caption{Position and velocity tracking and estimation performance of a static platform experiment using our BLSC. The tracking error remains small during the flight.}
\label{fig:static_yes}
\end{center}
	\begin{center}
		\includegraphics[trim =40mm 5mm 40mm 0mm, clip, width=1\textwidth]{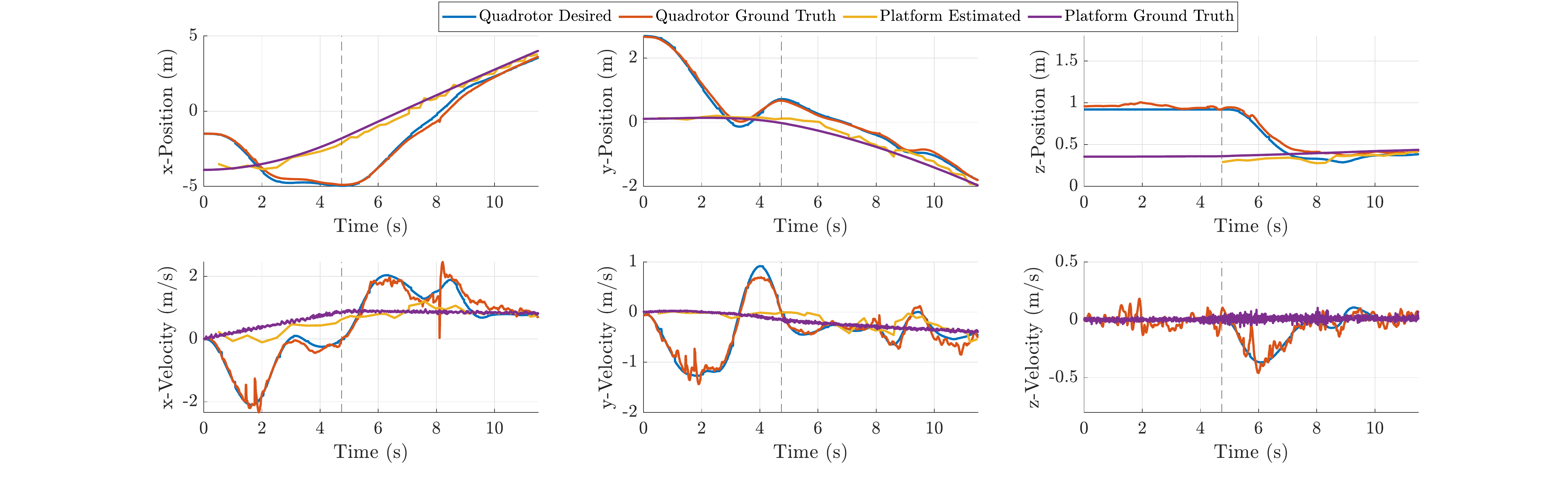}
		\vspace*{-.175in}
\caption{Position and velocity tracking and estimation performance of a moving platform experiment using our BLSC. The quadrotor quickly matches the UGV's velocity and is able to track successfully the planned trajectory.}
		\label{fig:moving_yes}
	\end{center}
\end{figure*}

\subsubsection{Results}
First, we tested a standard BLSC that does not take into account the turbulent wind generated by the leaf blowers, that is, the factors $\hat{f}$ and $K$ only consider the drag generated by the quadrotor's speed relative to the ground. Fig.~\ref{fig:static_no} shows the tracking and estimation performance obtained. The quadrotor starts at $\boldsymbol{p_q}=(-3.5, 2.0, 1.3)$m, and the platform is located at $\boldsymbol{p_p}=(1.1, -1.1, 0.8)$m. As expected, the tracking performance is poor, and the landing takes a long time: $27.2$s since the first tag detection, which occurs at $t_d=3.3$s (indicated with a vertical dashed line). Also note that the platform's estimation improves after $t_d$. In the video, the covariance ellipses for the tag's 2D position can be seen, and decrease abruptly in size at $t_d$.

Next, we used our BLSC with the same initial conditions as in Fig.~\ref{fig:static_no} to compare its improvement. The results are shown in Fig.~\ref{fig:static_yes}. It can be seen that the tracking is much better, and it only worsens slightly when the quadrotor is inside the wind field, after the time of the first tag detection $t=2.5$s. The landing time is just $3.2$s (measured since the first tag detection) even in challenging conditions, which is 8.5 times faster compared to the standard BLSC.


\subsection{Moving Platform Experiments}
\subsubsection{Experiment Setting}
To fully test our approach, we also performed landing experiments on a moving platform that is mounted on top of a dolly. The \textit{Clearpath Jackal} was used as the ground vehicle that tows the landing platform (see Fig.~\ref{fig:video-screenshot}) and carries two of the leaf blowers. This provided turbulent air at the landing pad even though the vehicles were not moving very quickly. The distance from the leaf blowers to the platform is such that this turbulent wind follows the same plot as Fig.~\ref{fig:windsp}.

\subsubsection{Results}
A standard BLSC was also tested first for this experiment setting. There was considerable tracking error and the quadrotor was not able to land on the platform before the vehicle arrived at the final point (see video linked in the Supplementary Material Section for details).
The results obtained using our BLSC are shown in Fig.~\ref{fig:moving_yes}. The quadrotor starts at $\boldsymbol{p_q}=(-1.5, 2.7, 0.9)$m, and the UGV starts at $\boldsymbol{p_p}=(-3.9, 0.1, 0.4)$m. When the maneuver begins, the UGV is commanded to move at 1m/s and rotate to its right at a rate of 4$^\circ$/s. The tag is detected at $t_d = 4.7$s, and the landing time is $6.8s$. Note that the quadrotor is at approximately $4m$ from the moving platform at $t_d$, a value larger than for the static experiment (1.5m).

\section{CONCLUSION}\label{s:conclusion}
This paper developed a boundary layer sliding controller to allow a quadrotor to fly in challenging conditions, and demonstrated its effectiveness by performing fast landing experiments. Future work includes incorporating adaptation to allow for more varied flight conditions, and landing on the back of a pickup truck driving outdoors, to test this work's approach in a more realistic setting. This will additionally require visual-inertial odometry (VIO) for onboard-only state estimation.

\section*{ACKNOWLEDGMENT}

Work supported by Ford Motor Company, and the authors would like to thank Michael Everett for his help configuring the UGV.

\balance




\bibliographystyle{IEEEtran}
\bibliography{IEEEabrv,biblio}

\end{document}